\documentclass[11pt]{article}
\usepackage{amsmath,amssymb,color}

\usepackage{amsfonts}

\makeatletter
\@addtoreset{equation}{section}
\makeatother

\newcommand{\hoch}[1]{$\, ^{#1}$}

\newcommand{\be}{\begin{equation}}
\newcommand{\ee}{\end{equation}}
\newcommand{\bea}{\setlength\arraycolsep{2pt} \begin{eqnarray}}
\newcommand{\eea}{\end{eqnarray}}
\newcommand{\nn}{\nonumber}

\def\ft#1#2{{\textstyle{\frac{\scriptstyle #1}{\scriptstyle #2} } }}
\def\fft#1#2{{\frac{#1}{#2}}}

\def\0{{\sst{(0)}}}
\def\1{{\sst{(1)}}}\def\2{{\sst{(2)}}}
 \def\3{{\sst{(3)}}}
\def\4{{\sst{(4)}}}
\def\5{{\sst{(5)}}}
\def\6{{\sst{(6)}}}
\def\7{{\sst{(7)}}}
\def\8{{\sst{(8)}}}
\def\sst#1{{\scriptscriptstyle #1}}
\def\oneone{\rlap 1\mkern4mu{\rm l}}
\def\ep{{\epsilon}}
\def\del{{\partial}}
\def\im{{{\rm i}}}
\def\cG{{{\cal G}}}
\def\cA{{{\cal A}}}
\def\cF{{{\cal F}}}
\def\cM{{{\cal M}}}

\def\cV{{{\cal V}}}
\def\cB{{{\cal B}}}
\def\cU{{{\cal U}}}

\def\tPi{{{\widetilde\Pi}}}

\def\tr{{\rm tr}}

\def\crampest{\medmuskip = 1mu plus 1mu minus 1mu}
\def\uncramp{\medmuskip = 4mu plus 2mu minus 4mu}

\begin{document}

\begin{flushright}
\hfill UPR-1254-T\ \ \ \ MIFPA-13-31
\end{flushright}

\vspace{25pt}
\begin{center}
{\Large {\bf Electrodynamics of Black Holes in STU Supergravity}}

\vspace{20pt}
{\Large M. Cveti\v c\hoch{1,2}, G.W. Gibbons\hoch{3}, C.N. Pope\hoch{3,4} 
     and Z.H. Saleem\hoch{1,5} }

\vspace{10pt}

\hoch{1}{\it Department of Physics and Astronomy,\\
 University of Pennsylvanian, Philadelphia, PA 19104, USA}

\vspace{10pt}

\hoch{2}{\it Center for Applied Mathematics and Theoretical Physics,\\
University of Maribor, Maribor, Slovenia}

\vspace{10pt}

\hoch{3}{\it DAMTP, Centre for Mathematical Sciences,
 Cambridge University,\\  Wilberforce Road, Cambridge CB3 OWA, UK}

\vspace{10pt}

\hoch{4}{\it George and Cynthia Woods Mitchell Institute for Fundamental
Physics and Astronomy,\\
Texas A\&M University, College Station, TX 77843-4242, USA}

\vspace{10pt}

\hoch{5}{\it National Center for Physics, Quaid-e-Azam university,\\ 
Shahdara Valley Road, Islamabad, Pakistan}

\vspace{30pt}

\underline{ABSTRACT}
\end{center}

External magnetic fields can probe the composite structure of black holes 
in string theory.  With this motivation we study magnetised four-charge 
black holes in the STU model, a consistent truncation of  maximally 
supersymmetric supergravity with four types of electromagnetic fields. 
We employ solution generating techniques to obtain Melvin backgrounds,  
and  black holes in these backgrounds.  For  an initially  electrically 
charged static black hole immersed in  magnetic fields,   we calculate 
the resultant angular momenta and analyse their  
global structure. Examples are given for which the ergoregion does not 
extend to infinity. We calculate magnetic moments and  gyromagnetic ratios  
via Larmor's formula. Our results are consistent with earlier special 
cases.  A scaling limit and associated subtracted geometry in a  single 
surviving magnetic field is shown to lift to $AdS_3\times S^2$. Magnetizing 
magnetically charged black holes give static solutions with conical 
singularities representing strings or struts holding the black holes 
against magnetic forces.  In some cases it is possible to balance these 
magnetic forces. 

\thispagestyle{empty}

\pagebreak

\tableofcontents
\addtocontents{toc}{\protect\setcounter{tocdepth}{2}}



\section{Introduction}

In recent work \cite{GMP,GPP} the structure and thermodynamics of 
charged and rotating black holes in Einstein-Maxwell theory 
immersed in an external  magnetic field were studied. The solutions 
were obtained by means of a solution-generating technique pioneered by
Ernst \cite{Ernst}, starting from a seed solution
with no external magnetic field present. By the well-known electric-magnetic 
duality  of Einstein-Maxwell theory,
one may just as easily consider the effects due to the 
immersion in an electric field of an electrically and magnetically charged
rotating black hole. The solution-generating property of
Einstein-Maxwell theory extends to ungauged supergravity theories,
and nowadays is best seen as part of the web of dualities 
at the heart of the modern synthesis of supergravity and superstring 
theory known as M-Theory. The supergravity theories of interest  contain
some number $k>1$ of generalised Maxwell fields, and black holes
may thus carry $k$ generalised electric and $k$ generalised  magnetic 
charges. Explicit solutions are available for all eight charges 
(four electric and four magnetic) \cite{chowcomp} 
 in a theory  often referred to as the STU
model, which is ${\cal N}=2$ supergravity coupled to 3 additional
vector multiplets.  (See \cite{cvetyoum,cclp} for the results for
just four charges.)  
If this is reduced to three dimensions on a Killing
symmetry, the bosonic sector of the 
resulting theory can be cast into the form of a scalar
sigma model with a global $O(4,4)$ symmetry coupled to gravity.  By
acting with an appropriate subgroup of $O(4,4)$ on the spatial reduction of
a four-charge black hole, one can introduce further parameters that
acquire the interpretation of describing external electric and/or magnetic
fields after lifting the solution back to four dimensions.
It is therefore of interest  to ask
how these black holes respond to being immersed  in a 
combination of the 4 possible electric and 4 possible
magnetic fields. 

Even in the Einstein-Maxwell case, the general situation is 
extremely complicated because a magnetic field   may  exert 
a torque  on a charged black hole  and cause
it to rotate, even if it was originally non-rotating and static.
Another potential complication is that applying, for instance,
a magnetic field to a magnetically charged black hole should cause it
to accelerate. In fact, in the case of Einstein-Maxwell theory,
it was found that a static magnetically charged but electrically
neutral Reissner-Nordstr\"om black hole seed remains static and 
non-accelerating upon magnetization,
but the metric then exhibits a conical singularity along the axis
of symmetry which represents a cosmic string whose  tension is  tuned
so as to  prevent acceleration.

It is of interest to explore these phenomena further when 
more than one electromagnetic field is present, since, as we shall
show in this paper, new features 
then arise. In view of the many
complications introduced  by rotation, we have decided to focus
in this paper just on static seed solutions.  

   The theory that we shall be
considering, the four-dimensional STU supergravity model, can arise as 
a 6-torus reduction of ten-dimensional type IIA (or heterotic)
supergravity to give ${\cal N}=8$ (${\cal N}=4$)  supergravity in 
four dimensions,
followed by a consistent truncation.  The truncation can be performed 
in a variety of different ways that are all equivalent under four-dimensional
U-duality, but which have different ten-dimensional interpretations, 
depending on how the four surviving gauge fields are selected.\footnote{A 
black hole solution of the STU model, specified by five independent charges, 
is a  generating solution for general black holes of the full  
${\cal N}=8$ (${\cal N}=4$)  supergravity theory with  28 electric and 
28 magnetic charges, which are obtained by acting on the generating solution 
with a subset of $U$-duality (or $S$ and $T$-duality)  transformations 
(See, for example, \cite{CH}).  The seed solution (in fact presented,
for symmetry reasons,  with eight charges rather than the minimal
set of five charges) has recently been constructed in \cite{chowcomp}.
The five-charge  static and BPS black holes were obtained 
in \cite{CYI} and \cite{CT}.} 

   In the
bulk of the paper we shall consider the STU model in the formulation
that was used in \cite{cvetyoum,cclp}, in which two of the gauge fields come
from the reduction of the  NS-NS  2-form potential of IIA supergravity, and
the other two are Kaluza-Klein vectors coming from the reduction of the
ten-dimensional metric.  The four-dimensional 
Reissner-Nordstr\"om black hole lifts to a pp-wave/NUT/NS1/NS5 
intersection in this description.  

In an alternative description of the STU model one of the gauge potentials
is the direct reduction of the Ramond-Ramond vector potential in the IIA theory, 
whilst the rest of the gauge potentials come from the Kaluza-Klein
reduction of the Ramond-Ramond 3-form potential.  In this description the 
Reissner-Nordstr\"om black hole lifts to a D0/D4/D4/D4 intersection in
ten dimensions.  (Alternatively, in   M-theory this configuration is obtained 
from a pp-wave/M5/M5/M5 intersection.)  We shall discuss the relation 
between the above  two 
formulations of the STU model in appendix B.

  There does not seem to be an ideal and succinct way of referring to the
two basic types of solution that we shall be discussing in this
paper.  In essence, we wish to consider the 4-field STU model generalisations
of two distinct magnetised Reissner-Nordstr\"om (RN) black holes:  
\bea
(1) && \hbox{Magnetised RN black hole carrying electric charge}
\nn\\
(2)&& \hbox{Magnetised RN black hole carrying magnetic charge}
\label{rn12}
\eea
Because of the duality complexions of the four field strengths in the 
STU model, the 4-field generalisation of solution 1 above actually has two
field strengths with electric charges and external magnetic fields, while
the other two field strengths have magnetic charges and external electric 
fields.  The situation is the opposite for the 4-field generalisation of
solution 2, in the sense that the first two field strengths now have 
magnetic charges and external magnetic fields, while the remaining two 
field strengths have electric charges and external electric fields.  

   In order to avoid a cumbersome description of these two types of solution
we shall sometimes for brevity 
refer to them as if we were working in a duality complexion where 
all four field strengths carried electric charges and external magnetic 
fields in the generalisations of solution 1, and all four fields carried
magnetic charges and external magnetic fields in the generalisations 
of solution 2.\footnote{Unfortunately, as we discuss in appendix B, in
actuality the Lagrangian of the STU model in 
such a duality complexion would be extremely complicated and inconvenient.}
Thus, in summary, we shall refer to the STU model generalisations of
solution 1 as magnetised electric black holes, and
the generalisations of solution 2 as magnetised magnetic black holes.

   Before constructing the STU model generalisations of solutions 1 and 2 
above, we begin in section 3 by obtaining the STU model generalisation of
the pure Melvin magnetic universe.  This can be constructed by starting
from Minkowski spacetime as the seed solution, acting with
the appropriate $O(4,4)$ global symmetries after reduction to three 
dimensions, and then lifting back to four dimensions.  We also show
how it can alternatively be constructing in a manner that generalises
a procedure described in \cite{gibhei}, as an analytic 
continuation of a limiting form of the four-charge static black hole solutions
in the STU model.   

In section 4 we construct
the STU model generalisations of solution 1 above (the magnetised
Reissner-Nordstr\"om solution with electric charge).  The solutions
we obtain have four independent charges and four independent external fields.
As one might 
anticipate from the
results of \cite{GMP}, they are in general stationary, even though we
again start from a static seed solution (the four-charge black holes of
the STU model).  This is because the external fields exert torques on
the charges carried by the black hole.  We show that it is in fact possible,
by choosing the charges and the external fields appropriately, to
balance  the torques and thereby obtain a static black hole solution.
We also examine the asymptotic structure of the metrics in the general
case, showing that generically there is an ergoregion extending out
to infinity, close to the axis.  We discuss the conditions on the charges
and magnetic fields under which the metrics become asymptotically static
at infinity, with no ergoregions.

For many purposes it is helpful to focus
on the simplifications that result by taking a {\it near horizon limit}
of the black hole metrics. In \cite{Cvetic:2012tr}, 
the resulting {\it subtracted geometry} \cite{CLI,CLII,virmani} 
was obtained by taking a suitable scaling limit.
In section 5, we apply this idea the to the metrics considered in this 
paper.

  In section 6
we turn to the STU model generalisation of solution 2 above (the magnetised 
Reissner-Nordstr\"om solution with magnetic charge).  
The solutions, which are all static, again have four independent charges
and four independent external fields.  
As one might have anticipated on the basis of the results
in \cite{GMP}, the metrics in general have a conical singularity
on the axis, corresponding to a delta-function tension holding
the black hole in place.   Interestingly, this may be eliminated by
imposing an appropriate condition on the charges and magnetic fields.
This can be interpreted as being due to a cancellation of the forces
associated with the individual charges and fields.

In addition to the concrete results obtained above, 
we include in the appendices  some more technical material
in which the explicit calculations and comparisons 
between different formalisms used are described in more  detail
than is done in the body of the text.  

\section{The STU Model and its Black Holes}
     
In this paper we shall be studying some properties of black holes in
the four-dimensional STU supergravity theory, which comprises ${\cal N}=2$ 
supergravity coupled to three vector multiplets.  The Lagrangian for 
the bosonic sector of the STU model, in the notation
of \cite{cclp}, is 
\bea
{\cal L}_4 &=& R\, {*\oneone} - \ft12 {*d\varphi_i}\wedge d\varphi_i 
   - \ft12 e^{2\varphi_i}\, {*d\chi_i}\wedge d\chi_i - \ft12 e^{-\varphi_1}\,
\Big( e^{\varphi_2-\varphi_3}\, {* F_{\2 1}}\wedge F_{\2 1}\nn\\
&& + e^{\varphi_2+\varphi_3}\, {*  F_{\2 2}}\wedge F_{\2 2}
   + e^{-\varphi_2 + \varphi_3}\, {* \cF_\2^1 }\wedge \cF_\2^1 + 
     e^{-\varphi_2 -\varphi_3}\, {*\cF_\2^2}\wedge \cF_\2^2\Big)\nn\\
&& + \chi_1\, (F_{\2 1}\wedge \cF_\2^1 + 
                  F_{\2 2}\wedge \cF_\2^2)\,,
\label{d4lag}
\eea
where the index $i$ labelling the dilatons $\varphi_i$ and axions $\chi_i$
ranges over $1\le i \le 3$.  The four field strengths can be written in 
terms of potentials as
\bea
F_{\2 1} &=& d A_{\1 1} - \chi_2\, d\cA_\1^2\,,\nn\\
F_{\2 2} &=& d A_{\1 2} + \chi_2\, d\cA_\1^1 - 
    \chi_3\, d A_{\1 1} +
      \chi_2\, \chi_3\, d\cA_\1^2\,,\nn\\
\cF_\2^1 &=& d\cA_\1^1 + \chi_3\, d \cA_\1^2\,,\nn\\
\cF_\2^2 &=& d\cA_\1^2\,.
\eea
Note that (\ref{d4lag}) could be obtained by reducing the six-dimensional
bosonic string action on $S^1\times S^1$, and then dualising the 2-form
potential $A_\2$ to the axion that is called $\chi_1$ here.

   Four-charge rotating black hole solutions in the STU theory were
constructed in \cite{cvetyoum}.  We shall use the conventions and notation
of \cite{cclp}, in which the metric for the four-charge black holes is 
given by
\be
ds_4^2 = -\fft{\rho^2-2mr}{W}\, (dt+ \cB_\1)^2 + 
    W\, \Big(\fft{dr^2}{\Delta} + d\theta^2 + 
   \fft{\Delta\, \sin^2\theta\, d\phi^2}{\rho^2-2mr}\Big)\,. \label{4dmetric}
\ee
where
\bea
\Delta &=& r^2 -2m r + a^2,\qquad \rho^2 = r^2 + a^2 \cos^2\theta\,,\nn\\
\cB_\1 &=& \fft{2ma \sin^2\theta (r\Pi_c -
                      (r-2m)\Pi_s)}{(\rho^2 -2mr)}\,d\phi\,,\nn\\
W^2&=&r_1\, r_2\, r_3\, r_4 + a^4 \cos^4\theta  \nn\\
&&+ a^2
\cos^2\theta \, [2r^2 + 2m r \sum_{i=1}^4 s_i^2 
+ 8m^2 \Pi_s\, \Pi_c - 4m^2(2
\Pi_s^2 + \sum_{i=1}^4 \Pi_s^i)]\,,
\eea
$r_i=r+2m s_i^2$, $s_i=\sinh\delta_i$, $c_i=\cosh\delta_i$, and 
$\Pi_c=c_1 c_2 c_3 c_4$ and $\Pi_s = s_1 s_2 s_3 s_4$.  We also define
\be
\Pi_s^i =s_i^{-1}\, \Pi_s\,,\qquad \Pi_c^i = c_i^{-1}\, \Pi_c\,.
\ee
The expressions for
the gauge 
potentials, axions and dilatons can be found \cite{cclp}.  

  The mass physical $M$, angular momentum $J$, charges $Q_i$ and
dipole moments $\mu_i$ were calculated in \cite{cvetyoum}.  In the notation
and conventions of \cite{cclp} that we are using here, they are given by
\bea
M &=& \ft14 m\sum_{i=1}^4 (c_i^2 + s_i^2)\,,\qquad J= m a\, (\Pi_c-\Pi_s)\,,\nn
\\
Q_i &=& 2 m s_i\, c_i\,,\qquad 
 \mu_i = 2m a\, (s_i\, \Pi_c^i - c_i\, \Pi_s^i)\,.\label{charges}
\eea

   In standard Maxwell electrodynamics, the magnetic moment of a particle
of mass $M$ and angular momentum $J$ carrying a charge $Q$ is given by
$\mu = {\bf g} J Q/(2M)$, where ${\bf g}$ is the gyromagnetic ratio.  
Generically,
for the four-charge black holes in the STU model, we can expect a relation
of the form
\be
\mu_i = \fft{J}{2M}\, \sum_{j=1}^4\, {\bf g}_{ij}\, Q_j\,.
\ee
From the quantities (\ref{charges}) given above it is not possible, in the
absence of additional criteria, to derive a unique form for the
``gyromagnetic matrix'' ${\bf g}_{ij}$.  However, if we impose the 
additional requirements that it be a symmetric matrix, and furthermore
that it exhibit the same symmetries as the metric under permutation of
the four charge parameters $\delta_i$, then we are led to the 
following result:
\bea
i=j:\qquad\qquad&&   {
 \bf g}_{ii} = \fft1{2c_i^2}\, \sum_{k=1}^4(c_k^2 + s_k^2)\,,\nn\\
i\ne j:\qquad\qquad && {\bf g}_{ij}= -\fft{\Pi_s}{6 c_i c_j s_i s_j}\, 
\fft{\sum_{k=1}^4( c_k^2+s_k^2)}{\Pi_c-\Pi_s}\,.\label{gyro}
\eea

   In special cases the expression (\ref{gyro}) for the gyromagnetic
ratio reduces to previously-known results.  For example, if we consider the
single-charge case where $\delta_2=\delta_3=\delta_4=0$ then we obtain
the ``Kaluza-Klein'' result \cite{gibwil,Yaza2}
\be
{\bf g} = {\bf g}_{11} = 2-\tanh^2\delta_1\,.
\ee
If two or more of the charges are non-zero, the gyromagnetic matrix has
off-diagonal components.  
If we take all four charges to be equal, then
\be
{\bf g}_{ij}=  \fft{2(c^2+s^2)}{c^2}\,,\ i=j\,,\qquad 
{\bf g}_{ij}= -\fft{2 s^2}{3 c^2}\,,\ i\ne j\,,
\ee
and so with $Q_i=Q$ we have ${\bf g}_{ij} Q_j= 2Q$, implying the standard
result  \cite{Carter} that ${\bf g}=2$ for the Kerr-Newman black hole.

In the case of two non-zero equal charges, say, $Q_1=Q_2=Q$ and $Q_3=Q_4=0$, we obtain the following nonzero gyromagnetic matrix coefficients:
 \be {\bf g}_{11}={\bf g}_{22}=2
 \,, \   {\bf g}_{33}={\bf g}_{44}=2\,c^2\, , \ {\bf g}_{34}={\bf g}_{43}=-\fft{2}{3}s^2\, .\ee
 Thus,    ${\bf g}_{1j}Q_j={\bf g}_{2j}Q_j=2Q$ which implies  ${\bf g}=2$ for $Q$.

 In the case of  three non-zero equal charges, say, $Q_1=Q_2=Q_3=Q$ and $Q_4=0$, we get the following nonzero gyromagnetic matrix coefficients:
 \bea && {\bf g}_{11}={\bf g}_{22}={\bf g}_{33}=2+\tanh^2\delta \, , \ {\bf g}_{44}=3c^2-1
 \,, \\ \nonumber  &&{\bf g}_{i4}={\bf g}_{4i}= -\fft{1}{3}\tanh^2\delta (2+\tanh^2\delta) \, ,\ i=1,2,3 \, .\eea
In this case ${\bf g}_{ij}Q_j=(2+\tanh^2\delta)Q$ for $i=1,2,3$, and thus ${\bf g}=2+\tanh^2\delta$. Furthermore, even though $Q_4=0$, a nonzero $\mu_4$ is induced,  since ${\bf g}_{4j}Q_j=-\tanh^2\delta(2+\tanh^2\delta)$ and thus   ${\bf g}_4= -\tanh^2\delta(2+\tanh^2\delta)$.

Another explicit example can be obtained with pair-wise equal charges, say, $Q_1=Q_3$ and $Q_2=Q_3$. In this case
the pair-wise equal magnetic moments $\mu_1=\mu_3$ and $\mu_2=\mu_4$ are related to the pair-wise equal charges as:
\be \mu_I=\fft{J}{2M}\sum_{J=1}^2 {\bf G}_{IJ} Q_J\, , \quad I=1,2\, , 
\ee
 where the coefficients of the gyromagnetic matrix ${\bf G}$ are 
\be {\bf G}_{11}=\fft{2(3c_1^2-2+2c_2^2)}{3c_1^2} \, , \ {\bf G}_{22}=\fft{2(3c_1^2-2+2c_2^2)}{3c_2^2}\, , \ {\bf G}_{12}={\bf G}_{21}=-\fft{4}{3}\fft{s_1s_2}{c_1c_2}\,  
\ee 

The matrix ${\bf G}$ has eigenvalues $2$ and $2+\fft{4}{3}(\tanh^2\delta_1+\tanh^2\delta_2)$.

\section{Pure Melvin-type Solution in the STU Model}

  Later in the paper, we shall be constructing solutions in the
STU model describing four-charge black holes immersed in external
magnetic fields.  These solutions will, under appropriate circumstances, 
be asymptotic to the STU model generalisations of the Melvin universe of
Einstein-Maxwell theory.  It is useful, therefore, 
first to consider the simpler
case of these pure Melvin-type solutions, where there is no black hole but
just the external magnetic fields.  (To be precise, as explained in the
introduction, when we use the expression ``external magnetic fields'' we
mean that the fields numbered 1 and 3 carry external electric fields, while
those numbered 2 and 4 carry external magnetic fields.)  The STU model in the
conventions we are using is given in appendix A.  Melvin-type
solutions can be found using the results presented in appendix A, starting
from a purely Minkowski seed solution. They
can also be read off from the expressions for magnetised black holes 
presented in section 3, by setting the black hole mass and charges to zero.
Thus the
metric is given by (\ref{4metric}) and (\ref{Hdef}) with $\omega=0$ and
\be
\Delta = \prod_{i=1}^4 \Delta_i\,,\qquad
\Delta_i = 1 +  \beta_i^2\, r^2\,\sin^2\theta\,,\label{Deltas}
\ee
and so
\be
ds_4^2 =\sqrt{\Delta}\, (- dt^2 + 
 dr^2 + r^2\, d\theta^2)  +
  \fft1{\sqrt{\Delta}}\, r^2\, \sin^2\theta\, d\phi^2\,.
\ee
Note that here, and throughout the rest of the paper, we use the notation that
\be
\beta_i= \ft12 B_i\,,
\ee
where $B_i$ is the physical asymptotic strength of the $i$'th  
field on the symmetry axis at large distance.  This is done in order to avoid
many cumbersome factors of $\ft12$ and powers of $\ft12$ in subsequent
formulae. In the pure Melvin case under discussion here, where there is
no black hole, the field strengths are in fact constant along the axis. 

   The scalar fields are given by
\be
e^{2\varphi_1}= \fft{\Delta_1 \Delta_3}{\Delta_2\Delta_4}\,,
\qquad
e^{2\varphi_2}= \fft{\Delta_2\Delta_3}{\Delta_1\Delta_4}\,,
\qquad
e^{2\varphi_3}=\fft{\Delta_1\Delta_2}{\Delta_3\Delta_4}\,,\label{melvinscalars}
\ee
with the axions all vanishing.
The four electromagnetic potentials
$\{A_{\1 1}, A_{\1 2}, \cA_\1^1,\cA_\1^2\}$ are given by
\bea
A_{\1 1} &=& -2\beta_1\, r\, \cos\theta\, dt\,,\qquad
\cA_\1^1 = -2\beta_3\, r\, \cos\theta\, dt\,,\nn\\
&&\nn\\
A_{\1 2}&=& \fft{\beta_2\, r^2\, \sin^2\theta}{\Delta_2}d\phi\,,\qquad
\cA_\1^2= \fft{\beta_4\, r^2\, \sin^2\theta}{\Delta_4}d\phi\,.
\label{melvingaugefields}
\eea

  In terms of cylindrical coordinates $(\rho,z)$ defined by $\rho=r\sin\theta$
and $z=r\cos\theta$, we have 
\be
ds_4^2 = \sqrt{\Delta}\, (-dt^2 +d\rho^2+dz^2) + \fft{\rho^2}{\sqrt{\Delta}}
\, d\phi^2\label{melvin2}
\ee
with $\Delta_i$ in (\ref{Deltas}) now given by $\Delta_i=1+\beta_i^2\, \rho^2$.
Making the further coordinate transformations to $x=\rho\, \cos\phi$ and 
$y= \rho\, \sin\phi$, the metric near the axis approaches Minkowski
spacetime $ds_4^2\rightarrow -dt^2 + dx^2 + dy^2 + dz^2$, and near the 
axis the field strengths approach
\be
F_{\2 1}\rightarrow B_1\, dt\wedge dz\,,\quad
F_{\2 2}\rightarrow B_2\, dx\wedge dy\,,\quad
\cF_\2^1\rightarrow B_3\, dt\wedge dz\,,\quad
\cF_\2^2\rightarrow B_4\, dx\wedge dy\,.
\ee
Thus, as mentioned above, the electric and magnetic field strengths have
magnitude $B_i$ on the axis for all values of $z$, in this pure Melvin case.

   It is interesting to note that the 4-field Melvin solution can be obtained
instead by means of a limiting procedure and analytic continuation from the
four-charge static black hole solution in the STU model, generalising the procedure
described in \cite{gibhei} for the Melvin solution in the Einstein-Maxwell
theory.  The four-charge black hole metric, which can be read off from
the magnetised black holes in section 3 by sending the magnetic fields 
$B_i$ to zero, is given by
\be
ds^2= -\fft{r(r-2m)}{\sqrt{r_1 r_2 r_3 r_4}}\, dt^2 +
  \sqrt{r_1 r_2 r_3 r_4}\, \Big[\fft{dr^2}{r(r-2m)} + d\theta^2 
  + \sin^2\theta\, d\phi^2\Big]\,,
\ee
where $r_i= r + 2m s_i^2$.  We then write the 2-sphere metric in the
form $d\theta^2 + \sin^2\theta\, d\phi^2= 4 (1+|\zeta|^2)^{-2}\, 
d\zeta d\bar\zeta$, where $\zeta= \tan\ft12\theta\, e^{\im \phi}$,
and perform the scalings
\be
r=\tilde r\, \lambda^{-1}\,,\qquad
t =\tilde t\, \lambda\,,\qquad m=\tilde m\, \lambda^{-3}\,,\qquad
s_i=\tilde s_i\, \lambda\,,\qquad \zeta= \tilde\zeta\, \lambda\,.
\ee
Sending $\lambda\rightarrow 0$ gives the metric
\be
ds^2 = \fft{2\tilde m\tilde r }{
\sqrt{\tilde r_1\tilde r_2\tilde r_3\tilde r_4}} \, d\tilde t^2 
+\sqrt{\tilde r_1\tilde r_2\tilde r_3\tilde r_4} \Big(
-\fft{d\tilde r^2}{\tilde m\tilde r} + 4 d\tilde\zeta d\bar{\tilde\zeta}\Big)
\,.
\ee
Defining 
\be
\tilde r =-\ft12 \tilde m\, \rho^2\,,\qquad \tilde \zeta=x+\im y\,,
\ee
and taking
\be
x=\ft12\im\,  \hat t\,,\qquad y=\ft12 z\,,\qquad
\tilde t= \fft{\im}{\tilde m}\, \tilde\phi\,,\qquad
\tilde s_i= \fft{\im}{2\beta_i}\,,
\qquad \tilde m= 2\sqrt{\beta_1\beta_2\beta_3\beta_4}\,,
\ee
we obtain the 4-field Melvin metric
\be
ds^2 = \sqrt{\Delta}\, (-d\hat t^2 + d\rho^2 + dz^2) +
\fft{\rho^2}{\sqrt{\Delta}}\,
d\tilde\phi^2\,,
\ee
where $\Delta=\prod_i\Delta_i$ with $\Delta_i=1+\beta_i^2\, \rho^2$.  We
see that this metric coincides with (\ref{melvin2}), after a minor 
change of notation.  Applying the same scalings and analytic continuations to
the scalar fields and gauge fields in the four-charge black hole solutions, 
one reproduces the results given in (\ref{melvinscalars}) and
(\ref{melvingaugefields}).

\section{Magnetised Electrically Charged Black Holes}

   Here, we consider the magnetisation of the four-charge solution of the
STU model that reduces, when the charges are set equal, to the magnetisation
of the electrically-charged Reissner-Nordstr\"om solution (i.e. it reduces 
to solution 1 in (\ref{rn12})).  Using the
notation and conventions of \cite{cclp}, this is achieved when the
field strengths numbered 1 and 3 carry magnetic charges, while the field
strengths numbered 2 and 4 carry electric charges.  In order to be able to
present the magnetised solution in the most compact way, we shall denote
the four charge parameters by $(q_1, q_2, q_3, q_4)$.

Applying the procedure described in appendix A, we find that
the metric is given by
\be
ds_4^2 = H\, [- r(r-2m) dt^2 + \fft{r_1 r_2 r_3 r_4}{r(r-2m)}\,dr^2 +
   r_1 r_2 r_3 r_4 d\theta^2\,]  
   + H^{-1}\, \sin^2\theta\, (d\phi -\omega dt)^2\,,
\label{4metric}
\ee
where
\be
r_i= r + 2m s_i^2\,,
\ee
and we shall use the notation $s_i=\sinh\delta_i$ and $c_i=\cosh\delta_i$.
The function $\omega$ is given by
\be
\omega = \sum_{i=1}^4 \Big[ -\fft{q_i\, \beta_i}{r_i} +
 \fft{q_i\,\Xi_i\,  [r_i+(r-2m)\cos^2\theta] r}{r_i}\Big]\,,
\label{omega}
\ee
where
\be
q_i= 2 m s_i c_i\,,\qquad 
\Xi_i= \fft{\beta_1\beta_2\beta_3\beta_4}{\beta_i}\,,\qquad 
\beta_i=\ft12 B_i\,,
\ee
and $B_i$ denotes the external magnetic field strengths for each of the 
four gauge fields.  Finally, the function $H$ is given in this case by
\be
H= \fft{\sqrt{\Delta}}{\sqrt{r_1 r_2 r_3 r_4}}\,,\label{Hdef}
\ee
where
\bea
\Delta &=& 1 + \sum_i\fft{\beta_i^2 r_1 r_2 r_3 r_4}{r_i^2}\, \sin^2\theta +
  2[\beta_3\beta_4 q_1 q_2+\cdots]\cos^2\theta
+[\beta_3^2\,\beta_4^2 \,R_1^2\, R_2^2 +\cdots]\nn\\
&& - 2 (\prod_j \beta_j r_j)\,  \sum_i\fft{q_i^2}{r_i^2}\, 
   \sin^2\theta\cos^2\theta 
 +[2\beta_2\beta_3\beta_4^2 q_2 q_3 \, R_1^2
   +\cdots]\cos^2\theta  +\prod_i \beta_i^2 \,R_i^2 \nn\\
&& + r_1 r_2 r_3 r_4\, 
\sum_i \fft{\Xi_i^2 \, R_i^2}{
    r_i^2}\, \sin^2\theta 
+
 [2\beta_1\beta_2\beta_3^2\beta_4^2 q_3 q_4\,
 R_1^2 \,R_2^2
  +\cdots]\cos^2\theta\,,
\eea
and we have defined 
\be
R_i^2 = r_i^2 \, \sin^2\theta + q_i^2\, \cos^2\theta\,.
\ee
Note that in each of the square-bracketed terms, the ellipses denote all
the analogous terms that arise by taking all inequivalent permutations
of the indices 1, 2, 3 and 4.

   The periodicity $\Delta\phi$ of the azimuthal coordinate $\phi$ 
is determined by the requirement that there should be no conical 
singularity at the north and south poles of the sphere.  Since 
$\Delta$ is an even function of $\cos\theta$, the requirements at the
north and the south poles are identical, and they imply that $\phi$ should
have period given by
\be
\Delta\phi = 2\pi\alpha\,,\qquad
\alpha=\Big( 1 + [\beta_1\beta_2 q_3 q_4 +\cdots] +
                      \prod_i \beta_i q_i\Big)\,,\label{phiperiod}
\ee
where the ellipses in the square brackets 
represent the five additional terms that follow from the
indicated term by taking all inequivalent permutations of the labels 1, 2, 3
and 4.

    The physical charges carried by the four gauge fields can be calculated 
easily using the expressions in appendix \ref{emcharges}.
The non-zero ones are $(P_1,Q_2,P_3,Q_4)$.  
For the sake of uniformity we shall change the notation and call these
$(\widetilde Q_1,\widetilde Q_2,\widetilde Q_3,\widetilde Q_4)$ 
respectively.  They turn out to be given by
\be
\widetilde Q_i = 
\fft{(q_i-\beta_i^2 q_1 q_2 q_3 q_4/q_i)}{\alpha}\, \fft{\Delta\phi}{2\pi}\,,
\ee
where $\alpha$ is defined in (\ref{phiperiod}). 
We therefore have
\be
\widetilde Q_i= q_i -\fft{\beta_i^2 \, q_1\, q_2\, q_3\, q_4}{q_i}\,.
\ee

   The solutions for the gauge potentials are given by 
\bea
A_{\1 1}&=& \beta_1\, r(r-2m)\cos\theta\, \Big[\fft1{r_1} -\fft{1}{r_2}
           -\fft1{r_3} -\fft1{r_4}\Big]\, dt + \sigma_1\, (d\phi-\omega dt)\,,
\nn\\
A_{\1 2} &=& \Big[-\fft{q_2}{r_2} + \sum_{i=1,3,4}
  \fft{r\, q_i\, \beta_1\beta_3\beta_4\, [r_i+ (r-2m)\cos^2\theta]}{
            \beta_i\, r_i}\Big]\, dt + \sigma_2\, (d\phi-\omega dt)\,,\nn\\
\cA_\1^1&=& \beta_3\, r(r-2m)\cos\theta\, \Big[\fft1{r_3} -\fft{1}{r_1}
           -\fft1{r_2} -\fft1{r_4}\Big]\, dt + \sigma_3\, (d\phi-\omega dt)\,,
\nn\\
\cA_\1^2 &=& \Big[-\fft{q_4}{r_4} + \sum_{i=1}^3 
  \fft{r\, q_i\, \beta_1\beta_2\beta_3\, [r_i+ (r-2m)\cos^2\theta]}{
            \beta_i\, r_i}\Big]\, dt + \sigma_4\, (d\phi-\omega dt)\,,
\eea
where $\sigma_i\equiv  \tilde\sigma_i/\Delta$, and
\bea
\tilde\sigma_1 &=& -q_1\cos\theta + \beta_1\cos\theta \Big[\fft{\beta_2}{r_2}\, 
(r_1 r_3 r_4 q_2\sin^2\theta - q_1 q_3 q_4 r_2 \cos^2\theta) +\cdots \Big] -
(\beta_3\beta_4 q_2  +\cdots)R_1^2 \cos\theta \nn\\
&& -\beta_1^2 \cos\theta \Big[ \fft{\beta_3\beta_4}{r_2}\, 
(r_1 r_3 r_3 q_2\sin^2\theta - q_1 q_3 q_4 r_2 \cos^2\theta) R_2^2+\cdots\Big] 
+\beta_1^3(\beta_2 q_2 R_3^2 R_4^2+\cdots)\cos\theta
\nn\\
&&+\fft{\beta_1\beta_2\beta_3\beta_4\cos\theta}{r_1} 
 (r_2 r_3 r_4 q_1 \sin^2\theta -q_2 q_3 q_4 r_1 \cos^2\theta) R_1^2
+\beta_1^3 \beta_2\beta_3 \beta_4 q_1 \cos\theta R_2^2 R_3^2 R_4^2\,,\\
\tilde\sigma_2 &=& \fft{\beta_2 r_1 r_3 r_4}{r_2}\sin^2\theta +
  (\beta_1 q_3 q_4+\cdots) \cos^2\theta 
+\beta_2(\beta_1^2 R_3^2 R_4^2+\cdots) +
2\beta_2(\beta_3\beta_4 q_3 q_4 R_1^2+\cdots)\cos^2\theta\nn\\
&&+ q_2[\beta_1^2(\beta_3 q_3 R_4^2 +\beta_4 q_4 R_3^2)+\cdots]\cos^2\theta
+ 4\beta_1 \beta_3 \beta_4 q_2 q_1 q_3 q_4 \cos^4\theta \nn\\
&&-
 \fft{\beta_1\beta_3\beta_4 q_2^2 r_1 r_3 r_4}{r_2}\sin^2\theta\cos^2\theta
-\beta_1\beta_3\beta_4\,r_2\,
\Big(\fft{q_1^2 r_3 r_4}{r_1} +\cdots\Big)\sin^2\theta\cos^2\theta\nn\\
&&
+\beta_1 \beta_3\beta_4(\beta_3\beta_4 q_3 q_4 R_1^2+\cdots) R_2^2\cos^2\theta 
 +\beta_2 r_2\Big[\fft{\beta_3^2 \beta_4^2 r_3 r_4}{r_1}\, R_1^4+\cdots\Big] 
\sin^2\theta \nn\\
&&+ 2\beta_1\beta_2\beta_3\beta_4 q_2(\beta_1 q_1 R_3^2 R_4^2 +
\cdots)\cos^2\theta + \beta_2\beta_1^2 \beta_3^2\beta_4^2 R_1^2 R_2^2 
R_3^2 R_4^2\,,\\
\tilde\sigma_3 &=&(\tilde\sigma_1\hbox{ with } 1\leftrightarrow 3)\,,\\
\tilde\sigma_4 &=& (\tilde\sigma_2 \hbox{ with } 2\leftrightarrow 4)\,.
\eea
When ellipses occur within a bracketed expression, they
denote the two additional terms obtained by
cycling the three index values taken from the set $\{1,2,3,4\}$ 
that are not equal to $i$.

   The axions and dilatons are given by 
\be
\chi_i = \fft{Z_i\cos\theta}{Y_i}\,,\qquad 
e^{2\varphi_i}= \fft{Y_i^2}{\Delta\, r_1 r_2 r_3 r_4}\,,\qquad i=1,2,3\,,
\ee
where
\bea
Z_1 &=& r_2 r_4[(\beta_1 q_3 + \beta_3 q_1) +
 \beta_2\beta_4 (\beta_1 q_1 R_3^2+ \beta_3 q_3 R_1^2)] \nn\\
&&\qquad -
   r_1 r_3[(\beta_2 q_4 + \beta_4 q_2) 
  +\beta_1\beta_3 (\beta_2 q_2 R_4^2+\beta_4 q_4 R_2^2)]\,,\\
Y_1 &=& r_1 r_3(1+2 \beta_1\beta_3 q_2 q_4 \cos^2\theta +
\beta_1^2 \beta_3^2 R_2^2 R_4^2) \nn\\
&& \qquad + r_2 r_4(\beta_1^2 R_3^2+\beta_3^2 R_1^2 
  + 2\beta_1\beta_3 q_1 q_3 \cos^2\theta)\,,\\
(Z_2,Y_2) &=& (-Z_1,Y_1) \hbox{ with } 1\leftrightarrow 2\,,\\
(Z_3,Y_3) &=& (Z_1,Y_1) \hbox{ with } 2 \leftrightarrow 3\,.
\eea

\subsection{Angular momentum}

   The angular momentum can be calculated using the standard procedure
developed by Wald.  The details of this calculation, and, in particular, 
the evaluation of the angular momentum in terms of the quantities in the
dimensionally-reduced three-dimensional theory, are given in \cite{GPP}.
A subtlety in the calculation concerns the different boundary conditions that
arise depending upon whether a gauge field carries an electric charge or a
magnetic charge.  If the charges were all electric, then the conserved 
angular momentum corresponding to the Killing vector 
$\xi=\del/\del\tilde\phi$, where $\tilde\phi= \phi/\alpha$ is the 
rescaled azimuthal coordinate that has period $2\pi$ and $\alpha$ 
is defined in (\ref{phiperiod}), would be \cite{GPP}
\be
J=\fft{\alpha}{16\pi}\, \int_{S^2} d(\chi_4 + \sigma_i\, \psi_i)\wedge d\phi =
   \fft{(\Delta\phi)^2}{32\pi^2}\, 
\Big[\chi_4 + \sigma_i\, \psi_i\Big]_{\theta=0}^{\theta=\pi}\,.
\ee
As discussed in \cite{GPP}, this expression is invariant under the
$U(1)^4$ abelian gauge transformations of the four gauge potentials that
preserve the condition that the Lie derivatives of the gauge potentials
with respect to the azimuthal Killing vector $\del/\del\phi$ vanish.

In our case, however, the fields $A_{\1 1}$ and $\cA_\1^1$ carry magnetic,
rather than electric, charges.  A simple way to evaluate the angular
momentum is to perform dualisations of these two fields.  Although rather
involved in the four-dimensional theory itself, the dualisations can be
easily implemented in the reduced three-dimensional theory, since then
they amount to exchanging the roles of the $\sigma_i$ and $\psi_i$ axions
for the fields in question. As can be seen from (\ref{kkdual}), since the
the Kaluza-Klein vector $\bar\cB_\1$ must be invariant under duality it
follows that the required duality transformations require also sending
\be
\chi_4+\sigma_i\, \psi_i \longrightarrow \chi_4 +\sigma_i\, \psi_i 
 - \sigma_1\, \psi_1 - \sigma_3\, \psi_3\,.
\ee
The conserved angular momentum for the four-charge black holes is therefore
given by
\be
J=
   \fft{(\Delta\phi)^2}{32\pi^2}\, \Big[\chi_4
    + \sigma_2\, \psi_2 + \sigma_4\, \psi_4\Big]_{\theta=0}^{\theta=\pi}\,.
\ee
Evaluating this, we find
\be
J =\ft12[\beta_1\, q_2 q_3 q_4+\cdots] +
  \ft12 q_1 q_2 q_3 q_4\, [q_1 \beta_2 \beta_3\beta_4 + \cdots]\,,
\label{Jtrue}
\ee
where the ellipses in each case denote the additional three symmetry-related
terms.

\subsection{Pairwise equal charges}

A considerable simplification arises in the function $\Delta$ 
if we set the fields pairwise equal,
so that
\be
B_3=B_1\,,\qquad B_4=B_2\,,\qquad \delta_3=\delta_1\,,\qquad \delta_4=
\delta_2\,.
\ee
We then find that
\be
\Delta = \Big[1+ \sum_{i=1}^2 \beta_i^2 (
 r_i^2\sin^2\theta + q_i^2\cos^2\theta) + 
  4 \beta_1\beta_2 q_1 q_2 \cos^2\theta +
     \prod_{i=1}^2 \beta_i^2 (r_i^2\sin^2\theta + q_i^2\cos^2\theta)\Big]^2\,.
\ee

   With the fields set pairwise equal, i.e. 
$q_3=q_1$, $q_4=q_2$ and $\beta_3=\beta_1$
and $\beta_4=\beta_2$.  We then find
\bea
\cA_\1^2 &=& \Big[-\fft{q_2}{r_2} +
\beta_1^2 q_2\, r\, \Big(1+\fft{(r-2m)}{r_2}\, \cos^2\theta\Big) +
  2\beta_1\beta_2 q_1\, r\, \Big(1+\fft{(r-2m)}{r_1}\, \cos^2\theta\Big)
\Big]\, dt\nn\\
&& + \sigma_4\, (d\phi-\omega dt)\,,\nn\\
\cA_\1^1 &=& -\fft{2\beta_1 r(r-2m)}{r_2}\, \cos\theta\, dt
 + \sigma_3\, (d\phi-\omega dt)\,,
\eea
with analogous expressions for $A_{\1 1}$ and $\cA_\1^1$. The
fields $\sigma_3$ and $\sigma_4$ are given by
\bea
\sigma_3&=& -q_1\cos\theta\, (1-\beta_1^2 R_2^2)\, Y^{-1}\,,\nn\\
\sigma_4 &=& \Big[\beta_2 R_1^2 +
  2\beta_1 q_1 q_2 \cos^2\theta +
  \beta_1^2\beta_2 R_1^2 R_2^2 \Big]\, Y^{-1}\,,
\eea
where
\bea
R_i^2 &=& r_i^2\, \sin^2\theta + q_i^2\, \cos^2\theta\,,\nn\\
Y&=& 1+ \beta_1^2 R_2^2 + \beta_2^2 R_1^2 + 4 \beta_1\beta_2 q_1 q_2
\cos^2\theta + \beta_1^2 \beta_2^2 R_1^2 R_2^2\,.
\eea

A different specialisation arises if we instead reverse the sign of the
fields $B_3$ and $B_4$ before the pairwise identification, in other words,
if we set
\be
B_3=-B_1\,,\qquad B_4=-B_2\,,\qquad \delta_3=\delta_1\,,\qquad \delta_4=
\delta_2\,.
\ee
Now, the function $\Delta$ becomes instead
\crampest
\bea
\Delta&=& [1+ 2\beta_1 q_2 \cos\theta + 
   \beta_1^2 (r_1^2\sin^2\theta + q_1^2\cos^2\theta)]
      [1- 2\beta_1 q_2 \cos\theta + 
   \beta_1^2 (r_1^2\sin^2\theta + q_1^2\cos^2\theta)]\times\nn\\
&&
[1+ 2\beta_2 q_1 \cos\theta + 
   \beta_2^2 (r_2^2\sin^2\theta + q_2^2\cos^2\theta)]
      [1- 2\beta_2 q_1 \cos\theta +
   \beta_2^2 (r_2^2\sin^2\theta + q_2^2\cos^2\theta)]\,.
\eea
\uncramp
Note that in this case the function $\omega$ now vanishes, and so the metric
is purely static.  In fact it is not hard to show that all the possible
ways of making $\omega$ vanish involve making one or another of the following
choices  
\bea
(1)\qquad &&q_i=q_j\,,\qquad q_k=q_\ell\,,\qquad 
 B_i=-B_j\,,\qquad B_k=-B_\ell\nn\\
(2)\qquad && q_i=q_j\,,\qquad q_k=-q_\ell\,,\qquad          
 B_i=-B_j\,,\qquad B_k=B_\ell\nn\\
(3)\qquad && q_i=-q_j\,,\qquad q_k=q_\ell\,,\qquad
 B_i=B_j\,,\qquad B_k=-B_\ell\nn\\
(4)\qquad && q_i=-q_j\,,\qquad q_k=-q_\ell\,,\qquad
 B_i=B_j\,,\qquad B_k=B_\ell\,,\label{staticcon}
\eea
where $i$, $j$, $k$ and $\ell$ are all different and are chosen from 1, 2, 3
and 4.  It can easily be seen that, as one would expect, the angular
momentum (\ref{Jtrue}) vanishes in all of these cases.

\subsection{Asymptotic structure and ergoregions}

   It was observed in \cite{GMP} that the metric component $g_{tt}$ in the
magnetised electrically charged Reissner-Nordstr\"om solution becomes
arbitrarily large and positive at large distances near to the $z$ axis,
thus indicating the presence of an ergoregion extending to infinity.  
Not surprisingly, the same is in general true in the STU model generalisations
of this solution that we are considering here.  Specifically, if we 
introduce cylindrical coordinates $\rho=r\, \sin\theta$ and 
$z=r\, \cos\theta$, then it is easily seen from (\ref{4metric}), (\ref{omega})
and (\ref{Hdef}) that to leading order in large $z$ and small $\rho$
we shall in general have
\be
g_{tt} \sim + z^2\, \rho^2\, \Big(\sum_i \beta_i\, \Xi_i\Big)^2\,,
\ee
and thus an ergoregion extending to infinity. The reason for this metric 
behaviour is that the function $\omega$ given in (\ref{omega}) has the
large-$z$ expansion
\be
\omega= 2z\, \sum_{i=1}^4 q_i\, \Xi_i  -2m \sum_{i=1}^4 q_i\, \Xi_i\, 
   (1+ s_i^2) + {\cal O}(\fft1{z})\,.\label{omegaz}
\ee

   The ergoregion is avoided if one imposes the condition $\sum_i q_i\, \Xi_i
=0$ on the charges and magnetic fields, i.e. if
\be 
\beta_1\beta_2\beta_3\beta_4 \sum_{i=1}^4 \fft{q_i}{\beta_i} =0\,.
\label{noergo}
\ee
One way to achieve this is if one (or more) of the four field strengths is
set to zero; for example, by taking $q_4=0$ and $\beta_4=0$.  Under these
circumstances the metric is still stationary, as opposed to static, but
is asymptotically non-rotating at infinity.  It can be seen from (\ref{Jtrue})
that the angular momentum also vanishes in such a case.

Clearly there are also more
general ways to satisfy (\ref{noergo}), where all four fields are 
non-vanishing.  If we assume that (\ref{noergo}) is satisfied then it follows
from (\ref{omegaz}) that the asymptotic metric near the axis is rotating 
with an angular velocity
\be
\Omega_\infty = 2m \sum_{i=1}^4 q_i\, \Xi_i\, s_i^2 = 4m^2 
\beta_1\beta_2\beta_3\beta_4\, \sum_{i=1}^4 \fft{\sinh^3\delta_i\,
\cosh\delta_i}{\beta_i}\,.\label{Omegainf}
\ee
It can also be seen from (\ref{omega}) that if (\ref{noergo}) holds then 
on the black hole horizon at
$r=2m$, the angular velocity will be
\be
\Omega_H = \sum_{i=1}^4 \fft{q_i\, \beta_i}{2m c_i^2} = 
  \sum_{i=1}^4 \beta_i\, \tanh\delta_i\,.
\ee

Note that in general, the angular momentum (\ref{Jtrue})
is non-vanishing if (\ref{noergo}) is satisfied.

Of course if any of the
conditions enumerated in (\ref{staticcon}) holds, then not merely
is (\ref{noergo}) satisfied but the metric is non-rotating everywhere, and
also $J=0$.

\section{Scaling Limit, and Lift to Five Dimensions}

   The scaling limits   of our 
  magnetised non-extremal black holes, which will be  parameterised by 
$\tilde m$, $\tilde \Pi _s $, $\tilde \Pi _c$   and  
${\tilde  \beta}_i$ ($i=1,\cdots ,4$), can be obtained by taking a specific
scaling limit \cite{Cvetic:2012tr} of the
magnetised electric black holes of section 3 parameterised by 
$m,\delta_i,\beta_i$ with  $\delta_1=\delta_2 =\delta_3$.  
After taking the limit, the solution can then be lifted
to five dimensions, where it can be seen to be AdS$_3\times S^2$.

   The limit can be implemented by setting $\delta_1=\delta_2 =\delta_3$  and making the scaling  \cite{Cvetic:2012tr}:
\bea
m&=&\tilde m\, \ep\,,\qquad r=\tilde r\, \ep\,,\qquad
    t=\tilde t\,\ep^{-1}\,,\qquad \beta_i = \tilde\beta_i\, \ep\,,\quad 
i=1,2,3,4\,,\nn \\
\sinh^2\delta_4&=& \fft{\tPi_s^2}{\tPi_c^2-\tPi_s^2}\,,\qquad
\sinh^2\delta_i = (\tPi_c^2-\tPi_s^2)^{1/3}\, \ep^{-4/3}\,,\quad
i=1,2,3\,,\label{scaling}
\eea
where $\ep$ is then sent to zero.  

The implementation of the scaling limit  (\ref{scaling}) gives
\be
(d\phi-\omega dt) \longrightarrow d\phi -
   (\tilde\beta_1+\tilde\beta_2+\tilde\beta_3) d\tilde t 
  -\fft{2\tilde m\tilde\beta_4\, \tPi_c \tPi_s}{
    (\tPi_c^2-\tPi_s^2) \tilde r + 2\tilde m \tPi_s^2}\, d\tilde t\,,
\ee
and
\be
\Delta\longrightarrow 1 + \fft{8 \tilde m^3 \tilde\beta_4^2 
   (\tPi_c^2-\tPi_s^2)^2 \sin^2\theta}{
          (\tPi_c^2-\tPi_s^2) \tilde r + 2\tilde m \tPi_s^2}\,.
\ee
The quantities $\tilde\beta_1$, $\tilde\beta_2$ and $\tilde \beta_3$ 
drop out completely in the scaling limit if we send $\phi\longrightarrow 
  \phi+(\tilde\beta_1+\tilde\beta_2+\tilde\beta_3) \tilde t$.  We shall
assume from now on that this redefinition has been performed.
Therefore, the  obtained  scaling limit 
 of  magnetised non-extremal black holes  
depend only on  {\it four} independent parameters: ${\tilde m}$, ${\tilde \Pi}_c$, ${\tilde \Pi}_s$ and ${\tilde \beta}_4$.  

In the case of vanishing magnetic fields, $\beta_i=0$ 
it was possible  \cite{Cvetic:2012tr} to identify the scaling limits with
the {\it subtracted geometry} \cite{CLII}  of a   non-extreme
black hole parameterised by $\tilde m,\tilde \delta_i$. In that case
we have    
$\tilde \Pi _s \equiv \Pi_{i=1}^4\sinh{\tilde \delta}_i$  
and $\tilde \Pi_c\equiv\Pi_{i=1}^4\cosh{\tilde \delta}_i$, 
determined by (\ref{scaling}).  In our case we have no 
independent derivation of a subtracted geometry and so no unique
identification of $\tilde \delta_i$ is possible.

   The lifting of the subtracted geometry solution to five dimensions is given by
\be
ds_5^2 = e^{\varphi_1}\, ds_4^2 + e^{-2\varphi_1}\, (dz + \cA_\1^2)^2\,.
\ee
Applying the scaling limit (\ref{scaling}) here, together with 
$z=\tilde z\, \ep^{-1}$, we find that the five-dimensional metric
$ds_5^2$ scales as $\ep^{-2/3}$, and defining $ds_5^2 = \ep^{-2/3}\,
d\tilde s_5^2$ we have
\bea
d\tilde s_5^2 &=& 4\tilde m^2 (\tPi_c^2-\tPi_s^2)^{2/3}\, 
[d\theta^2 + \sin^2\theta\, (d\phi+\tilde\beta_4 d\tilde z)^2 ] \\
&&+\fft{(\tPi_c^2-\tPi_s^2) \tilde r + 2\tilde m \tPi_s^2}{2\tilde m
     (\tPi_c^2-\tPi_s^2)^{4/3}}\, d\tilde z^2
-  \fft{(\tPi_c^2-\tPi_s^2) \tilde r - 2\tilde m \tPi_c^2}{2\tilde m
     (\tPi_c^2-\tPi_s^2)^{4/3}}\, d\tilde t^2 \nn\\
&&-
  \fft{2\tPi_c \tPi_s}{(\tPi_c^2-\tPi_s^2)^{4/3}}\, d\tilde td\tilde z
 + \fft{4\tilde m^2 (\tPi_c^2-\tPi_s^2)^{2/3}}{\tilde r(\tilde r-2\tilde m)}\,
  d\tilde r^2\,.\nn
\eea
It can be seen that $\tilde\beta_4$ disappears from the five-dimensional
metric if we make the further coordinate redefinition
\be
\phi= \tilde\phi - \tilde\beta_4\, \tilde z\,,
\ee
This is a reflection of the fact that the magnetisation of the
four-dimensional gauge field associated with the Kaluza-Klein vector 
$\cA_\1^2$ of the five-dimensional reduction can be implemented (or, in
the above calculation, undone) by performing a rotation in the 
$(\phi,\tilde z)$ plane \footnote{The role of the specific  Melvin 
transformation as a coordinate transformation in the $(\phi, {\tilde z})$  
plane of the lifted geometry was  first observed for dilatonic black 
holes in \cite{Yazadjiev}.}. This transformation is related to a spectral 
flow in a dual conformal field theory interpretation of AdS$_3$ geometries. 

   Finally, if we define new coordinates $\rho$, $\sigma$ and $\tau$ by
\be
\tilde r = 2\tilde m\, \cosh^2\rho\,,\qquad
\tilde z = 2\im\, (2\tilde m)^{3/2}\, (\tPi_c \, \tau +\tPi_s\, \sigma)\,,
\qquad
\tilde t = 2\im\, (2\tilde m)^{3/2}\, (\tPi_c \, \sigma +\tPi_s\, \tau)\,,
\ee
the five-dimensional metric can be seen to become
\be
d\tilde s_5^2 = 16\tilde m^2(\tPi_c^2-\tPi_s^2)^{2/3}\, \Big[
  (-\cosh^2\rho\, d\tau^2 + d\rho^2 + \sinh^2\rho\, d\sigma^2)
+  \ft14(d\theta^2 +\sin^2\theta d\tilde\phi^2)
  \Big]\,,
\ee
which is the metric on AdS$_3\times S^2$\,.

\section{Magnetostatic Black Holes}

\subsection{Magnetised magnetically charged black holes}

   Here we exchange the roles of the electric and the magnetic charges in
the original four-charge seed solution.  
That is, the charges numbered 1 and
3 are now electric, while those numbered 2 and 4 are magnetic, in the
conventions of \cite{cclp}.  
(Some of the properties of the
resulting metrics were discussed previously in \cite{emp1,emp2,emp3,emp4}.)
We shall denote the four charge parameters by
$(p_1,p_2,p_3,p_4)$ in this case.  In the case that the charges are set equal,
the solution reduces to the magnetised magnetically-charge Reissner-Nordstr\"om
black hole.
Concretely, in the
original seed solution, reduced to three dimensions, we replace 
(\ref{phiseed}) and (\ref{axionseed}) by
\be
e^{2\varphi_1}= \fft{r_2\, r_4}{r_1\, r_3}\,,\qquad
e^{2\varphi_2}= \fft{r_1\, r_4}{r_2\, r_3}\,,\qquad
e^{2\varphi_3}= \fft{r_3\, r_4}{r_1\, r_2}\,,\qquad
e^{2\varphi_4}= r_1\, r_2\, r_3\, r_4\, \sin^4\theta\,,\label{phiseedmag}
\ee
and
\bea
\chi_1&=& 0\qquad \chi_2=0\,,\qquad \chi_3=0\,,\qquad \chi_4=0  \,,\nn\\
\sigma_1&=& 0\,,\qquad \sigma_2=p_2\cos\theta \,,\qquad
\sigma_3= 0 \,,\qquad\sigma_4=p_4\cos\theta \,,\nn\\
\psi_1&=& p_1\cos\theta\,,\qquad\psi_2 = 0\,,\qquad 
\psi_3= p_3\cos\theta\,,\qquad \psi_4 = 0\,,\label{axionseedmag}
\eea

   The metric of the magnetised solution will still be given by 
(\ref{4metric}), but now we have $\omega=0$ and
the function $\Delta$ in (\ref{Hdef}) is given by
\be
\Delta = \prod_{i=1}^4 \Delta_i\,,\qquad
\Delta_i= (1+  \beta_i p_i \cos\theta)^2 +
                     \beta_i^2\,r_i^2 \, 
           \sin^2\theta\,.
\ee

Because $\Delta$ is not an even function of $\cos\theta$ in this case,
the periodicity conditions on $\phi$ for the metric to be free of conical  
singularities are different at the north and south poles of the sphere.
Specifically, we find that the required periodicities are
\bea 
\theta=0:&&\qquad \Delta\phi= 2\pi \prod_i (1+\beta_i p_i)\,,\nn\\
\theta=\pi:&&\qquad \Delta\phi= 2\pi \prod_i (1-\beta_i p_i)\,.
\eea
The metric can be rendered free of conical singularities if the
charges and magnetic fields satisfy the ``no-force condition''
\be
\prod_i (1+\beta_i p_i) = \prod_i (1-\beta_i p_i)\,.\label{nocon}
\ee

   Using the expressions given in section \ref{emcharges}, we can calculate
the physical electric and magnetic charges carried by the four gauge fields.
In this case, the non-vanishing ones are $(Q_1,P_2,Q_3,P_4)$.  For the
sake of uniformity, we shall relabel these as 
$(\widetilde P_1,\widetilde P_2,\widetilde P_3,\widetilde P_4)$ respectively.
They turn out to be given by
\be
\widetilde P_i= \fft{p_i}{(1-\beta_i^2\, p_i^2)}\, \fft{\Delta\phi}{2\pi}\,.
\ee

   The electromagnetic potentials are given by
\bea
\hat A_{\1 1} &=& 
\Big[-\fft{p_1}{r_1} + \fft{2\beta_1 r(r-2m)\cos\theta}{r_1}
  -\fft{\beta_1^2 \, p_1\, [r_1^2 + r(r-2m)\, \cos^2\theta]}{r_1}\Big]\,
dt\,,\nn\\ 
\hat A_{\1 2} &=&\fft{p_2 \cos\theta + \beta_2 \, R_2^2}{\Delta_2}\, 
d\phi\,, \nn\\
\hat \cA_\1^1 &=&\Big[-\fft{p_3}{r_3} + \fft{2\beta_3 r(r-2m)\cos\theta}{r_3}
  -\fft{\beta_3^2 \, p_3\, [r_3^2 + r(r-2m)\, \cos^2\theta]}{r_3}\Big]\,
dt\,,\nn\\
\hat \cA_\1^2 &=& \fft{p_4 \cos\theta + \beta_4 \, R_4^2}{\Delta_4}\, 
d\phi\,,\label{magmagpot}
\eea
where $R_i^2 = r_i^2 \sin^2\theta +p_i^2\cos^2\theta$.
The scalar fields are given by
\bea
e^{2\varphi_1} &=& \fft{r_2\, r_4\, \Delta_1\, \Delta_3}{r_1\, r_3\, 
                     \Delta_2\,\Delta_4}\,,\qquad
e^{2\varphi_2} = \fft{r_1\, r_4\, \Delta_2\, \Delta_3}{r_2\, r_3\, 
                     \Delta_1\,\Delta_4}\,,\qquad
e^{2\varphi_3} = \fft{r_3\, r_4\, \Delta_1\, \Delta_2}{r_1\, r_2\, 
                     \Delta_3\,\Delta_4}\,,\nn\\
&&\nn\\
\chi_1 &=&0\,,\qquad \chi_2=0\,,\qquad \chi_3=0\,.\label{magmagscal}
\eea
\subsection{$SL(2,R)^4$ truncations of the sigma model}

   The three-dimensional scalar sigma model associated with the timelike 
or spacelike reduction of the four-dimensional STU model has an $O(4,4)$
global symmetry. The Lagrangian in the case of the timelike reduction
can be found in section 2.1 of \cite{cclp}.  The sixteen scalars comprise
the original three dilatons $(\varphi_1,\varphi_2, \varphi_3)$ and
three axions $(\chi_1,\chi_2,\chi_3)$ of the STU model; the Kaluza-Klein
scalar $\varphi_4$ and the axion $\chi_4$ dual to the Kaluza-Klein vector;
the four axions $\sigma_i$ coming from the direct dimensional reductions of
the four gauge potentials; and finally the four axions $\psi_i$ coming
from the dualisations of the four gauge potentials in the dimensionally-reduced
theory.

   If we restrict attention to purely static configurations then $\chi_4$
will vanish.  If we furthermore restrict to configurations where the axions
$(\chi_1,\chi_2,\chi_3)$ of the STU model vanish, then it can be seen from
the sigma-model Lagrangian in eqn (7) of \cite{cclp} that there are two
possible disjoint truncations of the remaining scalar fields for
which the vanishing of the four $\chi_i$ axions is consistent with their
equations of motion.\footnote{In \cite{cclp} a {\it timelike} 
reduction to three dimensions was performed.  Here, we are instead
reducing on the spacelike azimuthal Killing vector $\del/\del\phi$ rather
than the timelike Killing vector $\del/\del t$.  The formulae in 
\cite{cclp} can be repurposed to the spacelike reduction with very
straightforward modifications.  In particular, the three-dimensional
sigma-model Lagrangian in eqn (7) of \cite{cclp} will take the same form
in the case of the spacelike reduction, except that the kinetic terms 
for all the scalar fields will now have the standard negative sign 
appropriate to a Minkowski-signature theory.} 
  Specifically, we can have either
\be
\sigma_1=\psi_2=\sigma_3=\psi_4=0\label{trunc1}
\ee
or
\be
\psi_1=\sigma_2=\psi_3=\sigma_4=0\,.\label{trunc2}
\ee
In the truncation (\ref{trunc1}), if we define
\bea
&&u_1=\ft12(\varphi_1-\varphi_2+\varphi_3-\varphi_4)\,,\qquad
u_2=\ft12(-\varphi_1+\varphi_2+\varphi_3-\varphi_4)\,,\nn\\
&&
u_3=\ft12(\varphi_1+\varphi_2-\varphi_3-\varphi_4)\,,\quad
u_4=\ft12(-\varphi_1-\varphi_2-\varphi_3-\varphi_4)\,,\nn\\
&& \alpha_1=\psi_1\qquad\alpha_2=\sigma_2\,,\qquad \alpha_3=\psi_3\,,\qquad
  \alpha_4=\sigma_4\,,\label{trunc11}
\eea
then the three-dimensional sigma-model Lagrangian in equation (7) of 
\cite{cclp}, after the appropriate sign-changes because we are making
a spacelike reduction, becomes
\be
{\cal L}_{\rm scal}= \sum_{i=1}^4 \Big( -\ft12 (\del u_i)^2 - \ft12
    e^{2u_i}\, (\del\alpha_i)^2\Big)\,.\label{trunclag}
\ee
This can be recognised as describing the coset $[SL(2,R)/O(2)]^4$. 
Similarly, if we consider instead the truncations (\ref{trunc2}), then 
defining instead
\bea
&&u_1=\ft12(-\varphi_1+\varphi_2-\varphi_3-\varphi_4)\,,\quad
u_2=\ft12(\varphi_1-\varphi_2-\varphi_3-\varphi_4)\,,\nn\\
&&
u_3=\ft12(-\varphi_1-\varphi_2+\varphi_3-\varphi_4)\,,\quad
u_4=\ft12(\varphi_1+\varphi_2+\varphi_3-\varphi_4)\,,\nn\\
&& \alpha_1=\sigma_1\qquad\alpha_2=\psi_2\,,\qquad \alpha_3=\sigma_3\,,\qquad
  \alpha_4=\psi_4\label{trunc22}
\eea
gives again an $[SL(2,R)/O(2)]^4$ sigma model with Lagrangian 
(\ref{trunc11}). 

(Note that if we considered a timelike reduction on the coordinate $t$
rather than a spacelike reduction on the coordinate $\phi$, we would end up with
a Lagrangian like (\ref{trunclag}) except with a minus sign in front of the
exponential terms.  The coset in this case would be $[SL(2,R)/O(1,1)]^4$.)

  The truncation described by (\ref{trunc11}) corresponds to the case where
the gauge fields numbered 1 and 3 are purely electric, while those numbered
2 and 4 are purely magnetic. Since in this paper we always consider 
Melvin backgrounds where fields 1 and 3 carry external electric fields,
while 2 and 4 carry external magnetic fields, this means that we can remain
within the truncation if we additionally allow fields 1 and 3 to carry electric
charges, and fields 2 and 4 to carry magnetic charges.  This is precisely
the situation we considered in section 4, namely the STU model generalisations
of the magnetically-charged Reissner-Nordstr\"om black hole in an external
magnetic field.  It can indeed be seen from equations (\ref{magmagpot}) and
(\ref{magmagscal}), together with the staticity of the metric, that the
solutions fall within the class described by the truncation (\ref{trunc1})
and (\ref{trunc11}).

   By contrast, although the {\it charges} carried by the gauge fields in
the solutions in section 3 are compatible with the truncation described by
(\ref{trunc2}) and (\ref{trunc22}), the external fields are still appropriate
for the other truncation, (\ref{trunc1}) and (\ref{trunc11}), and so
the solutions in section 3 are not described by either of the truncated 
theories.  And indeed, the axions $\chi_i$ are non-zero and the metric is
not static.

\subsection{Multi-centre BPS black holes in external magnetic fields} 

  Returning to the truncation (\ref{trunc1}) and (\ref{trunc11}), we can
in fact use it to describe more general situations than the ``magnetised
magnetically charged'' black holes obtained in section 4.  In particular,
we can consider the case of multi-centre BPS black holes that are
then immersed in external fields,
provided that we align them all along a line so that we can apply the
``Melvinising'' transformation.  
For these
purposes, it is useful first to present the general expressions for the
transformations of the scalar fields under the ``Melvinising" transformations.
If we start with a seed solution for which the fields are denoted by
bars, then after the transformation we will have
\be
e^{u_i}= e^{\bar u_i}\, [(1+\beta_i\, \bar\alpha_i)^2 + \beta_i^2\, 
   e^{-2\bar u_i}]\,,\qquad
\alpha_i = \fft{\bar\alpha_i\, (1+\beta_i\, \bar\alpha_i) +
                       \beta_i\, e^{-2\bar u_i} }{
    (1+\beta_i\, \bar\alpha_i)^2 + \beta_i^2\, e^{-2\bar u_i} }\,.
\label{melvinising1}
\ee
In particular this means that the transformed function $\varphi_4$ that 
appears in the metric ansatz (\ref{metans}) is given by 
\be
e^{-2\varphi_4} = e^{-2\bar\varphi_4}\, 
   \prod_{i=1}^4 [(1+\beta_i\, \bar\alpha_i)^2 + \beta_i^2\, e^{-2\bar u_i}]\,.
\label{melvinising2}
\ee

  The multi-centre black holes in the STU model have metrics given by
\be
ds^2 = -\Big(\prod_{i=1}^4 H_i\Big)^{-1/2}\, dt^2 +
     \Big(\prod_{i=1}^4 H_i\Big)^{1/2}\, d\vec y^2\,,\label{multimet}
\ee
where the functions $H_i$ are harmonic in the 3-dimensional Euclidean space
with metric $d\vec y^2$.  For black holes aligned along an axis we can
conveniently use cylindrical coordinates in which
\be
d\vec y^2 = d\rho^2 + \rho^2 \, d\phi^2 + dz^2\,.\label{cylinder}
\ee
We shall take the harmonic functions to be given by
\be
H_i = 1 + \sum_a \fft{p^{(a)}_i}{\sqrt{\rho^2 + (z-z_a)^2}}\,,
\ee
where the charges $p^{(a)}_i$ are constants and the black holes are
located at the point $z_a$ on the $z$ axis. The metric is free of
conical singularities on the $z$ axis provided that $\phi$ has period $2\pi$.

    A field strength carrying
an electric charge is described by a potential of the form
\be
A_{\rm elec}^i = - H_i^{-1}\, dt\,,\label{elecpot}
\ee
while a field strength carrying a magnetic charge is described by a potential
of the form
\be
A^i_{\rm mag} = \sum_a \fft{p^{(a)}_i\, (z-z_a)}{
  \sqrt{\rho^2 + (z-z_a)^2}}\, d\phi\,.\label{magpot}
\ee
In our case, therefore, the potentials for fields 1 and 3 are of the form
(\ref{elecpot}), while those for fields 2 and 4 are of the form (\ref{magpot}).
In the dimensionally-reduced three-dimensional language this implies that
the axionic scalars $\alpha_i$ defined in (\ref{trunc11}) are all given in
this seed solution by
\be
\bar\alpha_i = \sum_a \fft{p^{(a)}_i\, (z-z_a)}{
  \sqrt{\rho^2 + (z-z_a)^2}} \,.\label{baralpha}
\ee
The dilatonic scalar fields $\vec \varphi=(\varphi_1,\varphi_2,\varphi_3)$ 
in this multi-centre seed solution are given
by 
\be
\vec\varphi= \ft12 \sum_i \epsilon_i\, \vec c_i\, \log H_i\,,
\ee
where 
\be
{\cal L}= \sqrt{-g}(R-\ft12(\del\vec\varphi)^2 -\ft14 \sum_i
e^{\vec c_i\cdot\vec\varphi}\, (F^i)^2)
\ee
and $\epsilon_i$ is $+1$ if field $i$ carries an electric charge and $-1$
if it carries a magnetic charge (see, for example, section 2.2 of \cite{lptx}).
Comparing the multi-centre metric given by (\ref{multimet}) and 
(\ref{cylinder}) with the reduction ansatz (\ref{metans}), we see that in the
multi-centre seed solution we shall have 
\be
e^{\bar\varphi_4}= \rho^2\, \Big(\prod_i H_i\Big)^{1/2}\,,
\ee
and hence, from (\ref{trunc11}), 
\be
e^{-\bar u_i} = \rho\, H_i\,.\label{baru}
\ee

   Applying the Melvinising transformations (\ref{melvinising1}), we 
obtain the ``magnetised magnetic'' multi-centre black holes with metrics
\be
ds^2 = e^{-\varphi_4}\,\rho^2\, \Big[-dt^2 + \Big(\prod_i H_i\Big)\, 
  (d\rho^2 + dz^2)\Big] + e^{\varphi_4}\, d\phi^2\,,
\ee
where $\varphi_4$ is given by (\ref{melvinising2}).  
Thus the metric is given by
\be
ds^2 = Z^{1/2}\, \Big[-\Big(\prod_{i=1}^4 H_i\Big)^{-1/2}\, dt^2 +
     \Big(\prod_{i=1}^4 H_i\Big)^{1/2}\,(d\rho^2 + dz^2 + 
   Z^{-1}\, \rho^2\, d\phi^2)\Big]\,,
\ee
where
\be
Z= \prod_{i=1}^4 [(1+\beta_i\, \bar\alpha_i)^2 + \beta_i^2\, e^{-2\bar u_i}]\,.
\ee

   There will in general now be conical singularities along the $z$ axis.  
This can be seen by looking at the form of the metric in the $(\rho,\phi))$ 
plane as $\rho$ tends to zero.  From (\ref{baralpha}) and (\ref{baru}) 
we see that
as $\rho$ tends to zero we shall have
\be
Z\rightarrow \prod_{i=1}^4 (1+\beta_i\, \bar\alpha_i)^2\,,\qquad
\bar\alpha_i\rightarrow \sum_a p^{(a)}_i\, \hbox{sign}(z-z_a)\,.
\ee
In the case of a single-centre black hole, the periodicity conditions 
for $\phi$ in order to avoid a conical singularity can be seen to reduce to 
those in equation (\ref{nocon}).

\section{Conclusions}

In string theory charged black holes may be regarded as having a composite 
structure arising from their microscopic description in terms of intersecting 
D-branes/M-branes.  This composite structure is reflected in the interactions 
of the black holes. In this paper we have demonstrated this by using as 
external probes the various types of magnetic fields capable of exciting 
each of these constituents.  We have found that the behaviour of black holes 
is indeed rather sensitive to which type of magnetic field is applied. By far 
the simplest case is that of Kaluza-Klein black holes, which are made up of 
a single constituent.  Somewhat counterintuitively it turns out that the 
Maxwell-Einstein case is the most complex, which may be ascribed to the 
fact that all the constituents and probes are turned on. 

Utilising the composite structure of charges and magnetic fields allows for 
a balance of different forces and torques and the taming of the extent of 
ergoregions. This work samples only a restricted subset of  static 
four-charge  generating black hole solutions.  We anticipate that further 
studies of rotating five-charge generating solutions will reveal an even 
richer structure.

\section*{Acknowledgements}

We are grateful to Yi Pang for discussions.
M.C.~is supported in part by DOE grant DE-SC0007901, the Fay R. and
Eugene L. Langberg Endowed Chair and the Slovenian Research Agency (ARRS).
C.N.P.~is supported in part by DOE grant DE-FG02-95ER40917.

\appendix

\section{The STU Model}

\subsection{Reduction of the STU model to $D=3$}

We can
``magnetise'' the black hole solutions by performing a spacelike reduction 
to three dimensions on the azimuthal Killing vector $\del/\del\phi$, and
then acting with the appropriate $O(4,4)$ transformations.  This is
analogous to the discussion in \cite{cclp}, except that there the
reduction was performed on the timelike Killing vector $\del/\del t$\footnote{One can also employ a seed solution with  analytically continued coordinates:  $t\to i\phi$ and $\phi \to it$, perform the reduction on the the timelike Killing vector of the analytically continued solution,  act  on it with the appropriate generators of $O(4,4)$ transformations defined in \cite{cclp}, and  finally,  analytically continue the  obtained solution back to original coordinates $(t,\phi)$.}.
Thus we make a standard Kaluza-Klein reduction with 
\be
ds_4^2 = e^{-\varphi_4}\, d\bar s_3^2 + e^{\varphi_4}\, 
   (d\phi + \bar\cB_\1)^2
\,,\label{metans}
\ee
and
\bea
A_{\1 1} &=& \bar A_{\1 1} + \sigma_1\, (d\phi + \bar\cB_\1)\,,\qquad
A_{\1 2} = \bar A_{\1 2} + \sigma_2\, (d\phi + \bar\cB_\1)\,,\nn\\
\cA_\1^1 &=& \bar\cA_\1^1 + \sigma_3\, (d\phi+\bar\cB_\1)\,,\qquad
\cA_\1^2 = \bar\cA_\1^2 + \sigma_4\, (d\phi+\bar\cB_\1)\,.
\eea
where, when necessary, we place bars on three-dimensional quantities in
order to distinguish them from four-dimensional ones.  Note that throughout,
we use the ordering $(A_{\1 1}, A_{\1 2}, \cA_\1^1,\cA_\1^2)$ for 
the potentials, with $\sigma_i$ being the axionic scalar coming from the
direct Kaluza-Klein reduction of the $i$'th potential, and so on.

  In three dimensions we then dualise 1-form potentials to scalars, in 
a fashion that is precisely analogous to the one described for the
timelike reduction in \cite{cclp}.  The upshot is that the
Kaluza-Klein 1-form $\bar\cB_\1$, whose field strength is 
$\bar \cG_\2=d\bar\cB_1$, 
is replaced by the axion $\chi_4$ with
\be
e^{2\varphi_4}\, {\bar *\bar \cG_\2} = d\chi_4 + \sigma_1\, d\psi_1 + 
    \sigma_2\, d\psi_2 +\sigma_3\, d\psi_3 +   \sigma_4\, d\psi_4\,,
\label{kkdual}
\ee
and the 1-form potentials in three dimensions 
coming from the reduction of the four 1-form potentials in four dimensions
are dualised to axions $\psi_i$ where
\bea
- e^{-\varphi_1 + \varphi_2 - \varphi_3 + \varphi_4}\, {\bar * \bar F_{\2 1}} 
&=& d\psi_1 + \chi_3\, d\psi_2 - \chi_1\, d\sigma_3 - \chi_1\,
\chi_3\, d\sigma_4\,,\nn\\
- e^{-\varphi_1 + \varphi_2 + \varphi_3 + \varphi_4}\, {\bar * \bar F_{\2 2}} 
&=& d\psi_2 -\chi_1\, d\sigma_4\,,\nn\\
- e^{-\varphi_1 - \varphi_2 + \varphi_3 + \varphi_4}\, {\bar * \bar\cF_{\2}^1} 
&=& d\psi_3 - \chi_2\, d\psi_2 - \chi_1\, d\sigma_1 + \chi_1\,
\chi_2\, d\sigma_4\,,\nn\\
- e^{-\varphi_1 - \varphi_2 - \varphi_3 + \varphi_4}\, {\bar * \bar\cF_{\2}^2} 
&=& d\psi_4 + \chi_2\, d\psi_1 - \chi_3\, d\psi_3 -
\chi_1\,d\sigma_2 + \chi_2\, \chi_3\, d\psi_2 \nn\\
&&- \chi_1\, \chi_2\, d\sigma_3 +
 \chi_1\, \chi_3\, d\sigma_1 - \chi_1\, \chi_2\, \chi_3\, d\sigma_4\,.
\eea

   The three-dimensional Lagrangian in terms of the dualised fields is
a non-linear sigma model coupled to gravity, and can be written as
\be
\bar {\cal L}_3= \sqrt{-\bar g}\, 
   [\bar R- \ft12 \tr(\del\cM^{-1}\, \del\cM)]\,,
\ee
where $\cM=\cV^T\cV$ and 
\be
\cV = e^{\ft12 \varphi_i H_i}\, \cU_\chi\, \cU_\sigma\, \cU_\psi\,.
\ee
Here
\bea
{\cal U}_\chi &=& e^{\chi_1\, E_{\chi_1}}\, e^{\chi_2\, E_{\chi_2}}\, 
                 e^{\chi_3\, E_{\chi_3}}\, e^{\chi_4\, E_{\chi_4}}\,,\nn\\
{\cal U}_\sigma &=& e^{\sigma_1\, E_{\sigma_1}}\, e^{\sigma_2\, 
  E_{\sigma_2}}\, e^{\sigma_3\, E_{\sigma_3}}\, 
                   e^{\sigma_4\, E_{\sigma_4}}\,,\nn\\
{\cal U}_\psi &=& e^{\psi_1\, E_{\psi_1}}\, e^{\psi_2\, 
  E_{\psi_2}}\, e^{\psi_3\, E_{\psi_3}}\, 
                   e^{\psi_4\, E_{\psi_4}}\,.
\eea
$H_i$ are the Cartan generators of $O(4,4)$, whilst $E_{\chi_i}$, 
$E_{\sigma_i}$ and $E_{\psi_i}$ are the positive-root generators.  (See
\cite{cclp} for a detailed description of the notation we are using here.)

\subsection{Magnetisation of the four-charge static black hole}

   The usual four-charge black hole carries 
electric charges (Q) and magnetic charges (P) in the order 
$(P_1,Q_2,P_3,Q_4)$, 
where 
we use our standard ordering ($A_{\1 1}, A_{\1 2}, \cA_\1^1, \cA_\1^2)$ for
the gauge fields.  The
static four-charge solution corresponds, in three dimensions, to
\bea
d\bar s_3^2 &=& [- r(r-2m) dt^2 + \fft{r_1 r_2 r_3 r_4}{r(r-2m)}\,dr^2 +
   r_1 r_2 r_3 r_4 d\theta^2\,]\, \sin^2\theta \,,\label{3metric}\nn\\
r_i&=& r + 2m s_i^2\,,
\eea
with
\be
e^{2\varphi_1}= \fft{r_1\, r_3}{r_2\, r_4}\,,\qquad 
e^{2\varphi_2}= \fft{r_2\, r_3}{r_1\, r_4}\,,\qquad
e^{2\varphi_3}= \fft{r_1\, r_2}{r_3\, r_4}\,,\qquad
e^{2\varphi_4}= r_1\, r_2\, r_3\, r_4\, \sin^4\theta\,,\label{phiseed}
\ee
and
\bea
\chi_1&=& 0\qquad \chi_2=0\,,\qquad \chi_3=0\,,\qquad \chi_4=0  \,,\nn\\
\sigma_1&=&-q_1 \cos\theta\,,\qquad \sigma_2=0\,,\qquad
\sigma_3=-q_3\cos\theta\,,\qquad\sigma_4=0\,,\nn\\
\psi_1&=&0\,,\qquad\psi_2 =q_2\cos\theta\,,\qquad 
\psi_3=0\,,\qquad \psi_4 = q_4\cos\theta\,,\label{axionseed}
\eea
The magnetisation of the four-charge solution can be implemented by transforming
the coset representative $\cM$ defined above according to
\be
\cM \longrightarrow S\cM S^T\,,
\ee
where $S$ is the $O(4,4)$ matrix
\be
S= \exp(\ft12 B_1\, E_{\psi_1} +\ft12 B_2 \, E_{\sigma_2} +
        \ft12 B_3\, E_{\psi_3} + \ft12 B_4\, E_{\sigma_4})\,,
\ee
with (constant) parameters $B_i$ being the asymptotic values of the magnetic
fields of the four field strengths.  One then retraces the steps of
dualisation and lifts the transformed solution back to four dimensions to
obtain the magnetised black hole.\footnote{We remind the
reader that, as discussed in the introduction, when we speak, for the
sake of brevity, of the ``magnetised electrically-charged black hole'' in the
STU model we mean the one for which the field strengths numbered 1 and 3 carry
magnetic charges and external electric fields, while those numbered 2 and 4
carry electric charges and external magnetic fields.}  The results are 
presented in section 3.

\subsection{Magnetic and electric charges}\label{emcharges}

    The physical charges can be calculated very easily using the 
dimensionally-reduced quantities in three dimensions.  Using the
standard ordering of the $U(1)$ gauge fields, namely 
$\{A_{\1 1}, A_{\1 2}, \cA_\1^1,\cA_\1^2\}$, the magnetic charges are
given by
\bea
P_1 &=& \fft1{4\pi} \int_{S^2} d A_{\1 1} =
    \fft1{4\pi} \int_{S^2}  d\sigma_1 \wedge d\phi
  =\fft{\Delta\phi}{4\pi} \, \Big[\sigma_1\Big]_{\theta=0}^{\theta=\pi}\,,\nn\\
&&\nn\\
P_2 &=& \fft1{4\pi} \int_{S^2} d A_{\1 2} =
    \fft1{4\pi} \int_{S^2}  d\sigma_2 \wedge d\phi
  =\fft{\Delta\phi}{4\pi} \, \Big[\sigma_2\Big]_{\theta=0}^{\theta=\pi}\,,\nn\\
&&\nn\\
P_3 &=& \fft1{4\pi} \int_{S^2} d \cA_\1^1 =
    \fft1{4\pi} \int_{S^2}  d\sigma_3 \wedge d\phi
  =\fft{\Delta\phi}{4\pi} \, \Big[\sigma_3\Big]_{\theta=0}^{\theta=\pi}\,,\nn\\
&&\nn\\
P_4 &=& \fft1{4\pi} \int_{S^2} d \cA_\1^2 =
    \fft1{4\pi} \int_{S^2}  d\sigma_4 \wedge d\phi
  =\fft{\Delta\phi}{4\pi} \, \Big[\sigma_4\Big]_{\theta=0}^{\theta=\pi}\,,
\eea
where $\Delta\phi$ is the period of the azimuthal coordinate $\phi$.

  The electric charges are given by integrating the equations of 
motion of the four fields $\{A_{\1 1}, A_{\1 2}, \cA_\1^1,\cA_\1^2\}$.
These give
\bea
Q_1 &=& \fft1{4\pi} 
 \int_{S^2} e^{-\varphi_1+\varphi_2-\varphi_3}\, {* F_{\2 1}}
+\cdots = \fft1{4\pi}\int_{S^2} d\psi_1\wedge d\phi=
  \fft{\Delta\phi}{4\pi} \, \Big[\psi_1\Big]_{\theta=0}^{\theta=\pi}\,,\nn\\
&&\nn\\
Q_2 &=& \fft1{4\pi} 
 \int_{S^2} e^{-\varphi_1+\varphi_2+\varphi_3}\, {*F_{\2 2}}
+\cdots = \fft1{4\pi}\int_{S^2} d\psi_2\wedge d\phi=
  \fft{\Delta\phi}{4\pi} \, \Big[\psi_2\Big]_{\theta=0}^{\theta=\pi}\,,\nn\\
&&\nn\\
Q_3&=&\fft1{4\pi} 
\int_{S^2} e^{-\varphi_1-\varphi_2+\varphi_3}\, {*\cF_\2^1}
+\cdots = \fft1{4\pi}\int_{S^2} d\psi_3\wedge d\phi=
  \fft{\Delta\phi}{4\pi} \, \Big[\psi_3\Big]_{\theta=0}^{\theta=\pi}\,,\nn\\
&&\nn\\ 
Q_4&=&\fft1{4\pi}
\int_{S^2} e^{-\varphi_1-\varphi_2-\varphi_3}\, {*\cF_\2^2}
+\cdots = \fft1{4\pi}\int_{S^2} d\psi_4\wedge d\phi=
  \fft{\Delta\phi}{4\pi} \, \Big[\psi_4\Big]_{\theta=0}^{\theta=\pi}\,.
\eea
(The ellipses here denote the additional terms in the equations of motion. 
In each case, the full set of terms conspire to give just the simple
expressions presented here in terms of the fields $\psi_i$.)

\section{STU Model in Other Duality Complexions}

   As we discussed before, the in the formulation \cite{cclp} that
we are using in this paper for the STU model, the usual four-charge black 
hole carries 
electric charges (Q) and magnetic charges (P) in the order 
$(P_1,Q_2,P_3,Q_4)$, where 
we use our standard ordering $(A_{\1 1}, A_{\1 2}, \cA_\1^1, \cA_1^2)$
for the gauge fields.  To
convert into the parameterisation used, for example, in \cite{virmani}, 
we need to dualise the potential $A_{\1 2}$ to $B_\1$, whose field strength
is the dual of $F_{\2 2}$.  
To do this, we start from the Lagrangian (\ref{d4lag}) and then 
add a Lagrange multiplier 
\be
{\cal L}_{LM} = 4 dB_\1 \wedge (F_{\2 2} -\chi_2 \, d\cA_\1^1 +
   \chi_3\, dA_{\1 1} -\chi_2 \chi_3\, d\cA_\1^2)\,,
\ee
treating $F_{\2 2}$ now as an independent field that
we solve for algebraically and substitute back into the total Lagrangian.
This leads to the dualised Lagrangian
\bea
\widetilde {\cal L}_4 &=& 
R\, {*\oneone} - \ft12 {*d\varphi_i}\wedge d\varphi_i
   - \ft12 e^{2\varphi_i}\, {*d\chi_i}\wedge d\chi_i 
-2 e^{\varphi_1-\varphi_2-\varphi_3}\, {*G_\2}\wedge G_\2
\nn\\
&&- 2e^{-\varphi_1}\,
\Big( e^{\varphi_2-\varphi_3}\, {* F_{\2 1}}\wedge F_{\2 1} 
   + e^{-\varphi_2 + \varphi_3}\, {* \cF_\2^1 }\wedge \cF_\2^1 +
     e^{-\varphi_2 -\varphi_3}\, {*\cF_\2^2}\wedge \cF_\2^2\Big)\nn\\
&& - 4 \chi_1\, F_{\2 1}\wedge \cF_\2^1 +4 dB_\1 \wedge
(\chi_3\, dA_{\1 1} -\chi_2\, d\cA_\1^1 -\chi_2\chi_3 \, d\cA_\1^2) \,,
\label{d4duallag}
\eea
where $G_\2=e^{-\varphi_1+\varphi_2+\varphi_3}\, {*F_{\2 2}}$, which is
written in terms of the potential $B_\1$ as
\be
G_\2 = dB_\1 - \chi_1\, d\cA_\1^2\,.
\ee

   If we now define
\bea
\widetilde\chi^1 &=& -\chi_1\,,\qquad \widetilde\chi^2=-\chi_2\,,\qquad
\widetilde\chi^3=\chi_3\,,\nn\\
h^I &=& f^{-1}\, e^{-\varphi_i}\,,\qquad 
  f^3= e^{-\varphi_1-\varphi_2-\varphi_3}\,,
\qquad G_{IJ}= \hbox{diag}\{(h^1)^{-2},(h^2)^{-2},(h^3)^{-2}\}\,,\nn\\
A_\1^{[0]} &=& \cA_\1^2\,,\qquad A_\1^{[1]}= B_\1\,,\qquad
  A_\1^{[2]}= A_{\1 1}\,,\qquad A_\1^{[3]}= \cA_\1^1\,,
\eea
then (\ref{d4duallag}) can be written precisely in the form of equation (A.20)
of \cite{virmani} (where the axions $\widetilde\chi_i$ are those
in \cite{virmani}):
\bea
\widetilde{\cal L} &=& R {*\oneone} - \ft12 G_{IJ}\, {*dh}^I\wedge dh^J -
  \ft32 f^{-2}\, {*df}\wedge df - \ft12 f^3\, {*F_\2^{[0]}}\wedge 
F_\2^{[0]}\nn\\
&&-\ft12 f^{-2}\, G_{IJ}\, {*d}\widetilde \chi^I \wedge d\widetilde\chi^J -
\ft12 f\, G_{IJ}\, ({*F_\2^{[I]}} + \widetilde\chi^I\, {*F_\2^{[0]}}) \wedge
(F_\2^{[J]} + \widetilde\chi^J\, F_\2^{[0]})\nn\\
&&+ \ft12 C_{IJK}\, \Big[\widetilde\chi^I\, F_\2^{[J]}\wedge F_\2^{[K]} +
 \widetilde\chi^I\, \widetilde\chi^J\, F_\2^{[0]} \wedge F_\2^{[K]} +
 \ft13 \widetilde\chi^I\, \widetilde\chi^J\,\widetilde\chi^K\,
  F_\2^{[0]}\wedge F_\2^{[0]}\Big]\,,
\eea
where $F_\2^{[\Lambda]}= dA_\1^{[\Lambda]}$ and $C_{IJK}=|\epsilon_{IJK}|$.
The charges carried by the four-charge black hole in \cite{cclp} will now be of 
the form 
($Q,P,P,P)$, where the fields are ordered ($A_\1^{[0]}$, $A_\1^{[1]}$,
$A_\1^{[2]}$, $A_\1^{[3]}$).

   Note that we can in principle perform a further transformation on the
Lagrangian (\ref{d4duallag}), and dualise the gauge potential $\cA_\1^2$ 
also.  This would result in a formulation where the standard four-charge
black hole in \cite{cclp} would be supported by four gauge fields that
all carried magnetic charge.  This dualisation can be achieved by adding 
a Lagrange
multiplier $4d\widetilde B_\1\wedge\cF_\2^2$ to (\ref{d4duallag}), and
then solving algebraically for $\cF_\2^2$ and substituting back into the
total Lagrangian. The equation for $\cF_\2^2$ is quite complicated, taking
the form
\be
\alpha\, {*\cF_\2^2} = H_\2 + \beta\, \cF_\2^2\,,\label{eqn1}
\ee
where
\bea
\alpha&=& e^{-\varphi_1 -\varphi_2-\varphi_3} +
\chi_1^2\, e^{\varphi_1 -\varphi_2-\varphi_3} +
\chi_2^2\, e^{-\varphi_1 +\varphi_2+\varphi_3} +
\chi_3^2\, e^{-\varphi_1 -\varphi_2+\varphi_3} \,,\qquad
\beta=2 \chi_1\, \chi_2\, \chi_3\,,\nn\\
H_\2 &=& d\widetilde B_\1 -\chi_2\,\chi_3\, dB_\1 
  -\chi_1\, \chi_3\, dA_{\1 1} + \chi_1\, \chi_2\, d\cA_\1^1 +
\chi_1\, e^{\varphi_1 -\varphi_2-\varphi_3}\, {*dB_\1} \nn\\
&&+
\chi_2\, e^{-\varphi_1 +\varphi_2+\varphi_3}  \, {*dA_{\1 1}} -
\chi_3\, e^{-\varphi_1 -\varphi_2+\varphi_3}\, {*d\cA_\1^1}\,.
\eea
Equation (\ref{eqn1}) can be solved for $\cF_\2^2$, giving
\be
\cF_\2^2 = -\, \fft{\alpha\, {*H_\2} +\beta\, H_\2}{\alpha^2+\beta^2}\,,
\ee
but the result seems to be rather too complicated to be useful.


\begin{thebibliography}{99}




\bibitem{GMP} G.W. Gibbons, A.H. Mujtaba and C.N. Pope,
{\it Ergoregions in magnetised black hole spacetimes},
Class.\ Quant.\ Grav.\  {\bf 30}, no. 12, 125008 (2013),
arXiv:1301.3927 [gr-qc].

\bibitem{GPP}G.W. Gibbons, Yi Pang  and C.N. Pope
{\it Thermodynamics of magnetised Kerr-Newman black holes},
arXiv:1310.3286 [hep-th].


\bibitem{Ernst}
F.J. Ernst, {\it Black holes in a magnetic universe,}
J. Math. Phys. {\bf 17}, 54 (1976).

\bibitem{chowcomp} D.D.K. Chow and G. Comp\` ere,
{\it Seed for general rotating non-extremal black holes of 
$N=8$ supergravity},
arXiv:1310.1925 [hep-th].

\bibitem{cvetyoum}
M. Cveti\v c and D. Youm,
 {\it Entropy of nonextreme charged rotating black holes in string theory,}
  Phys.\ Rev.\  D {\bf 54}, 2612 (1996),
hep-th/9603147.

  
\bibitem{cclp}Z.-W. Chong, M. Cveti\v c, H. L\"u and C.~N. Pope,
{\it Charged rotating black holes in four-dimensional gauged and ungauged
supergravities},
Nucl.\ Phys.\ B {\bf 717}, 246 (2005), hep-th/0411045.

\bibitem{CH}
M. Cveti\v c and C.M. Hull,
  {\it Black holes and U duality,}
  Nucl.\ Phys.\  B {\bf 480}, 296 (1996),
hep-th/9606193.

\bibitem{CYI}
M. Cveti\v c and D. Youm,
 {\it All the static spherically symmetric black holes of heterotic string 
on a six torus,}
Nucl.\ Phys.\ B {\bf 472}, 249 (1996),
hep-th/9512127.

\bibitem{CT}
 M. Cveti\v c and A.A. Tseytlin,
 {\it Solitonic strings and BPS saturated dyonic black holes,}
  Phys.\ Rev.\ D {\bf 53}, 5619 (1996),
  [Erratum-ibid.\ D {\bf 55}, 3907 (1997)],
hep-th/9512031.
  
 \bibitem{Cvetic:2012tr}
  M. Cveti\v c and G.W. Gibbons,
 {\it Conformal symmetry of a black hole as a scaling limit: A 
black hole in an asymptotically conical box,}
  JHEP {\bf 1207} (2012) 014, 
arXiv:1201.0601 [hep-th].
  
\bibitem{CLI}
  M. Cveti\v c and F. Larsen,
 {\it Conformal symmetry for general black holes,}
  arXiv:1106.3341 [hep-th].
  
\bibitem{CLII}
  M. Cveti\v c and F. Larsen,
 {\it Conformal symmetry for black holes in four dimensions,}
  JHEP {\bf 1209}, 076 (2012),
arXiv:1112.4846 [hep-th].

\bibitem{virmani}  A. Virmani,
{\it Subtracted geometry from Harrison transformations},
JHEP {\bf 1207}, 086 (2012),
arXiv:1203.5088 [hep-th].

\bibitem{gibhei} G.W. Gibbons and C.A.R. Herdeiro,
{\it The Melvin universe in Born-Infeld theory and other theories of 
nonlinear electrodynamics},
Class.\ Quant.\ Grav.\  {\bf 18}, 1677 (2001), hep-th/0101229.


\bibitem{Yazadjiev} 
  S.S. Yazadjiev,
 {\it Electrically charged dilaton black holes in external magnetic field,}
  Phys.\ Rev.\ D {\bf 87}, 084068 (2013), arXiv:1302.5530 [gr-qc].

\bibitem{gibwil} G.W. Gibbons and D.L. Wiltshire,
{\it Black holes in Kaluza-Klein theory},
  Annals Phys.\  {\bf 167}, 201 (1986), 
  [Erratum-ibid.\  {\bf 176}, 393 (1987)].

\bibitem{Yaza2} S.S. Yazadjiev,
{\it Thermodynamics of rotating charged dilaton black holes in an 
external magnetic field},
  Phys.\ Lett.\ B {\bf 723}, 411 (2013), arXiv:1304.5906 [gr-qc].

\bibitem{Carter} 
  B. Carter,
 {\it Global structure of the Kerr family of gravitational fields},
  Phys.\ Rev.\  {\bf 174}, 1559 (1968).

\bibitem{emp1}
  R. Emparan,
{\it Composite black holes in external fields},
  Nucl.\ Phys.\ B {\bf 490} (1997) 365, hep-th/9610170.

\bibitem{emp2}
 R. Emparan,
{\it Black diholes},
  Phys.\ Rev.\ D {\bf 61} (2000) 104009,
hep-th/9906160.

\bibitem{emp3}
R. Emparan and E. Teo,
{\it Macroscopic and microscopic description of black diholes},
  Nucl.\ Phys.\ B {\bf 610}, 190 (2001), hep-th/0104206.

\bibitem{emp4}
A. Chatrabhuti, R. Emparan and A. Taormina,
{\it Composite diholes and intersecting brane-anti-brane configurations 
in string/M theory},
  Nucl.\ Phys.\ B {\bf 573} (2000) 291, hep-th/9911007.

\bibitem{lptx} 
H. L\"u, C.N. Pope, T.A. Tran and K.W. Xu,
{\it Classification of p-branes, NUTs, waves and intersections},
  Nucl.\ Phys.\ B {\bf 511}, 98 (1998), hep-th/9708055.


\end{thebibliography}
\end{document}